\documentclass[aps, pre, floatfix, reprint]{revtex4-2}
\usepackage[colorlinks=true, allcolors=blue, bookmarks=false]{hyperref}
\usepackage{amsfonts, amsmath, amssymb}
\usepackage{graphicx}

\DeclareMathOperator{\Li}{Li}

\begin{document}

\title{Average weighted ratio of consecutive level spacings for \\ infinite-dimensional orthogonal random matrices}

\author{Wouter Buijsman} 

\email{buijsman@pks.mpg.de}

\affiliation{Max Planck Institute for the Physics of Complex Systems, N\"othnitzer Str. 38, 01187 Dresden, Germany}

\date{\today}

\begin{abstract}
The onset of quantum ergodicity is often quantified by the average ratio of consecutive level spacings. The reference values for ergodic quantum systems have been obtained numerically from the spectra of large but finite-dimensional random matrices. This work introduces a weighted ratio of consecutive level spacings, having the propery that the average can be computed numerically for random matrices of infinite dimension. A Painlev\'e differential equation is solved numerically in order to determine this average for infinite-dimensional orthogonal random matrices, thereby providing a reference value for ergodic quantum systems obeying time-reversal symmetry (provided that the time-reversal operator squares to the identity matrix). A Wigner surmise-inspired analytical calculation is found to yield a qualitatively accurate picture for the statistics of high-dimensional random matrices from each of the symmetry classes. For Poissonian level statistics, a significantly different average is found, indicating that the average weighted ratio of consecutive level spacings can be used as a probe for quantum ergodicity.
\end{abstract}

\maketitle

\section{Introduction}
Ergodic quantum systems show various universal statistical properties that can be described by random matrix theory~\cite{Porter65, Brody81, Guhr98, DAlessio16}. From a physics perspective, arguably the most well-studied example is the level spacing distribution~\cite{Mirlin99, Evers08, Weidemuller09, Abanin19}. Also from a random matrix theory point of view, the level spacing distribution received considerable attention both historically~\cite{Forrester03} and recently~\cite{Riser23, Tian24, Forrester25, Martinez-Azcona25, Shir25}. Observing the universal properties of a level spacing distribution requires a spectrum to have a uniform density, which can be accomplished by a (sometimes, ambiguous) procedure known as unfolding~\cite{Brody81, Gomez02, Fossion13, Berkovits21}. It was realized about two decades ago that the need for unfolding can be circumvented by considering the distribution of the ratio of consecutive level spacings rather than the level spacing distribution~\cite{Oganesyan07, Atas13}. The average ratio of consecutive level spacings is nowadays a widely used diagnostic for quantum ergodicity~\cite{Luitz15, Khemani19, Sa20, Suntajs20, Giraud22, Kawabata23, Hashimoto23}.

The reference values for the average ratio of consecutive level spacings for ergodic quantum systems have been obtained numerically from the spectra of large but finite-dimensional random matrices~\cite{Atas13}. Recently, a mix of analyical and numerical methods has been used to compute the average ratio of consecutive level spacings for infinite-dimensional unitary random matrices, thereby providing a reference value for ergodic quantum systems with broken time-reversal symmetry~\cite{Nishigaki24, Nishigaki25}. See Ref.~\cite{Dyson62} for a discussion on the relation between different symmetry classes and time-reversal symmetry, as well as the meaning of ``unitary" in this context. Seemingly, no corresponding results for infinite-dimensional orthogonal or symplectic random matrices are available. This work numerically computes an average weighted ratio of consecutive level spacings for infinite-dimensional orthogonal random matrices, modeling ergodic quantum systems obeying time-reversal symmetry (provided that the time-reversal operator squares to the identity matrix). This result could be useful in detecting small deviations from quantum ergodicity, which can be seen as a topic of timely interest~\cite{Kliczkowski24, Herrmann25, Pathak25}. Next, it shows a way of how level spacing distributions can be used to study ratios of consecutive level spacings. The corresponding averages for infinite-dimensional unitary and symplectic random matrices can also be computed, as pointed out in the outlook. 

The main result of this work is obtained in two steps. A Painlev\'e differential equation is solved numerically in order to obtain the level spacing distributions of infinite-dimensional orthogonal and symplectic random matrices~\cite{Jimbo80, Tracy93}. Next, the level spacing distribution of symplectic random matrices is interpreted as the distribution (up to a scaling by a factor $2$) of two consecutive level spacings of orthogonal random matrices through a rigorously established inter-relation~\cite{Mehta63, Forrester01, Forrester09}. As a side remark, it is mentioned that the distributions of multiple consecutive level spacings have been a topic of recent interest within the physics community~\cite{Bhosale23, Tekur24, Martinez-Azcona25}. From these distributions, the average $\langle q \rangle$ of the logarithmically weighted ratio $q = \ln(1 + r)$ of consecutive level spacings $r$ is computed numerically for infinite-dimensional orthogonal random matrices, giving $\langle q \rangle \approx 0.8100699350$. For Poissonian level statistics, a significantly different average $\langle q \rangle = 1$ is found, indicating that the weighted average ratio of consecutive level spacings can be used as a probe for quantum ergodicity.

The outline of this work is as follows. Sec.~\ref{sec: weight} introduces the weight function. Sec.~\ref{sec: surmise} discusses analytical calculations for low-dimensional random matrices, and establishes qualitative agreement with numerical results for random matrices of high dimension from each of the symmetry classes. Also, Poissonian level statistics are discussed here. Sec.~\ref{sec: infinite-average} introduces the numerical procedure, and computes the average weighted ratio of consecutive level spacings for infinite-dimensional orthogonal random matrices. Sec.~\ref{sec: finite-average} numerically studies the convergence towards this result for finite dimensions. Sec.~\ref{sec: conclusions} closes with conclusions and a brief outlook.

\section{Weight function} \label{sec: weight}
As outlined below, the numerical approach used in this work allows one to compute the level spacing distribution for orthogonal, unitary, and symplectic random matrices of infinite dimension. Let $s_n$ and $s_{n+1}$ denote two consecutive level spacings, and let $r_n = s_{n+1} / s_n$ denote their ratio. Notice that this definition is different from the sometimes used definition $\tilde{r} = \min(r, 1/r)$. The main focus of this work is on the average of the logarithmically weighted ratio of consecutive level spacings
\begin{align}
q_n 
& = \ln(1 + r_n) \\
& = \ln(s_n + s_{n+1}) - \ln(s_n). \label{eq: q}
\end{align}
This quantity does not depend on the mean level spacing: it does not change under the scalings $s_1 \to \lambda s_1$ and $s_2 \to \lambda s_2$ with $\lambda > 0$. The average of this quantity is given by the sum of the averages of $\ln(s_n + s_{n+1})$ and $-\ln(s_n)$. For Gaussian and circular random matrix ensembles, among others, the level spacing distribution of symplectic matrices is (up to a scaling by a factor $2$) given by the distribution of two consecutive level spacings of orthogonal matrices~\cite{Mehta63, Forrester01, Forrester09}. Away from the spectral edge, eigenvalue spacings of large-dimensional random matrices obey the same statistics under very mild conditions on the eigenvalue density~\cite{Kamien88, Brezin93, Deift04}. 

Let $P_\beta(s)$ and $P_\beta(r)$ denote the level spacing distribution and distribution of the ratio of consecutive level spacings, respectively. The Dyson index $\beta$ can take values $1$ (orthogonal random matrices), $2$ (unitary random matrices), or $4$ (symplectic random matrices). For orthogonal random matrices, the average $\langle q \rangle$ of $q$ is then given by
\begin{align}
\langle q \rangle 
& = \int_0^\infty \ln(1 + r) \, P_1 (r) \, dr 
\label{eq: q-ratio} \\
& = \int_0^\infty \bigg( \frac{1}{2} P_4 (s/2) - P_1 (s) \bigg) \ln(s) \, ds.
\label{eq: q-spacing}
\end{align}
Because of the relation~\eqref{eq: q-spacing} between $\langle q \rangle$ and the level spacing distributions $P_\beta (s)$, $\langle q \rangle$ is arguably a more natural quantity to consider than the average (unweighted) ratio of consecutive level spacings. Below, Eq.~\eqref{eq: q-spacing} is evaluated numerically for random matrices of infinite dimension.

\section{Wigner surmise} \label{sec: surmise}
The Wigner surmise provides a qualitatively accurate closed-form expression for the level spacing distributions of high-dimensional random matrices~\cite{Porter65}. It is obtained as the exact expression for two-dimensional Gaussian random matrices as
\begin{equation}
P_{W, \beta} (s) = A s^\beta e^{-B s^2},
\label{eq: Wigner-surmise}
\end{equation}
where the $\beta$-dependent parameters $A$ and $B$ fix the normalization and mean level spacing to unity. Substituting the level spacing distributions of Eq.~\eqref{eq: Wigner-surmise} in Eq.~\eqref{eq: q-spacing} and evaluating the integral gives an analytical estimate $\langle q \rangle_s$ of $\langle q \rangle$ for orthogonal random matrices given by
\begin{align}
\langle q \rangle_s
& = \frac{4}{3} + \ln \bigg( \frac{3 \pi}{16} \bigg) \\
& \approx 0.8041.
\end{align}
This value deviates by $0.74 \%$ from the value for infinite-dimensional orthogonal random matrices obtained below.

A more accurate analytical estimate results by considering the distribution of the ratios of consecutive level spacings. For Gaussian random matrices of dimension three, one can show that
\begin{equation}
P_{W, \beta} (r) = \frac{1}{Z} \frac{(r + r^2)^\beta}{(1 + r + r^2)^{1 + \frac{3}{2} \beta}},
\label{eq: Wigner-surmise-inspired}
\end{equation}
where the $\beta$-dependent parameter $Z$ fixes the normalization to unity~\cite{Atas13, Atas13-2}. This Wigner surmise-inspired expression has been found numerically to provide a qualitatively accurate description of the distribution of the ratio of consecutive level spacings for high-dimensional random matrices from each of the symmetry classes. By substituting Eq.~\eqref{eq: Wigner-surmise-inspired} in Eq.~\eqref{eq: q-ratio} and evaluating the integral, one obtains an analytical estimate $\langle q \rangle_r$ of $\langle q \rangle$ for orthogonal random matrices given by
\begin{align}
\langle q \rangle_r
& = \frac{3}{2} - \ln(2) \\
& \approx 0.8069.
\end{align}
This value deviates by $0.40 \%$ from the corresponding result for infinite-dimensional orthogonal random matrices. For unitary and symplectic random matrices, the corresponding estimates can be expressed analytically in terms of dilogarithms, and evaluate to $0.7624$ (unitary random matrices) and $0.7314$ (symplectic random matrices). Analytical expressions leading to these values can be found in the Appendix. The deviations from the averages for high-dimensional random matrices obtained below are $0.16 \%$ and $0.069 \%$, respectively.

For Poissonian (uncorrelated) level statistics with unit mean level spacing, the probability density for the level spacings is given by $e^{-s}$, irrespective of the matrix dimension. The distribution of $n$ consecutive level spacings is given by the $n$-fold convolution of this distribution,
\begin{equation}
P_{P,n} (s) = \frac{s^{n - 1} e^{-s}}{(n - 1)!},
\end{equation}
which is also known as the Erlang distribution. From Eq.~\eqref{eq: q}, it then follows that the equivalent average $\langle q \rangle_P$ for Poissonian level statistics is given by
\begin{align}
\langle q \rangle_P
& = \int_0^\infty \bigg(P_{P,2} (s) - P_{P,1} (s) \bigg) \ln(s) \, ds \\
& = 1.
\end{align}
This value is significantly different from the values obtained above, showing that $\langle q \rangle$ is able to discriminate between Poissonain and Wigner-Dyson level statistics. From this, it follows that $\langle q \rangle$ can be used as a probe for quantum ergodicity. An alternative approach to obtain the same result is to consider the distribution of consecutive level spacings
\begin{equation}
P_P (r) = \frac{1}{(1 + r)^2}
\end{equation}
\cite{Oganesyan07, Atas13}. Substituting this probability density in Eq.~\eqref{eq: q-ratio} and evaluating the integral again gives $\langle q \rangle_P = 1$.

\begin{figure}
\includegraphics[scale=0.9]{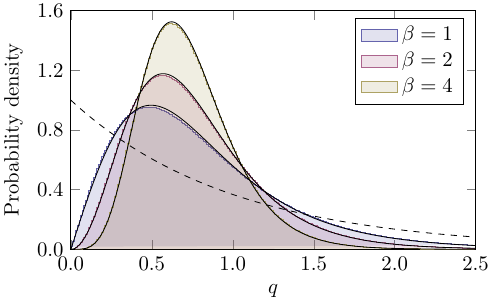}
\caption{The distributions of $q$ for high-dimensional orthogonal ($\beta = 1$), unitary ($\beta = 2$) and symplectic ($\beta =4$) random matrices. The histogram data has been obtained numerically from the spectra of $10000$ matrices sampled from circular ensembles of dimension $10,000$ ($\beta = 1,2$) or $5000$ ($\beta = 4$). The Wigner surmise-inspired probability distributions of Eq.~\eqref{eq: Wigner-surmise-inspired-q}, plotted in black, are qualitatively accurate. The distribution for Poissonian level statistics is shown by a dashed line.}
\label{fig: histograms}
\end{figure}

Eq.~\eqref{eq: Wigner-surmise-inspired} can be re-written as an analytical expression for the distribution of $q$ by a change of variables from $r$ to $q$. It follows that
\begin{equation}
P_{W, \beta} (q) =\frac{e^q}{Z} \frac{(r + r^2)^\beta}{(1 + r + r^2)^{1 + \frac{3}{2} \beta}}.
\label{eq: Wigner-surmise-inspired-q}
\end{equation}
Here, $Z$ is the same normalization constant as before, and $r = e^q - 1$. If $q$ is small compared to unity, this expression is proportional to $q^\beta$, while it is proportional to $\exp[-(1 + \beta) q]$ if $q$ is large compared to unity. For Poissonian level statistics, the corresponding expression is given by $e^{-q}$. Fig.~\ref{fig: histograms} compares this distribution for $\beta = 1$, $2$, and $4$ against numerical data for high-dimensional random matrices sampled from the circular ensembles, indicating that it provides a qualitatively accurate picture. The averages of $q$ for high-dimensional random matrices obtained from the numerical data are $0.7637(1)$ and $0.7319(4)$ for unitary and symplectic random matrices, respectively. Sec.~\ref{sec: finite-average} provides a more detailed discussion of the numerical approach. 

\section{Infinite-$\boldsymbol{N}$ average} \label{sec: infinite-average}
The Painlev\'e V differential equation provides a tool to numerically obtain the level spacing distributions of infinite-dimensional orthogonal, unitary, and symplectic random matrices~\cite{Jimbo80}. Reference~\cite{Edelman05} provides a practical introduction (including example code) of this method, which is loosely followed in the next two paragraphs. For infinite-dimensional orthogonal random matrices, it follows by using this method that
\begin{equation}
\langle q \rangle \approx 0.8100699350,
\end{equation}
which constitutes the main result of this work.

The Painlev\'e V differential equation, for numerical convenience written as a system of two coupled first-order differential equations, is given by
\begin{equation}
\frac{d}{dt} 
\begin{pmatrix}
\sigma \\
\sigma' 
\end{pmatrix}
=
\begin{pmatrix}
\sigma' \\
- \frac{2}{t} \sqrt{(\sigma - t \sigma') \, (t \sigma' - \sigma + (\sigma')^2)}
\label{eq: Painleve-V}
\end{pmatrix}.
\end{equation}
In order to obtain the level spacing distributions (where the mean level spacing is set to unity) for infinite-dimensional random matrices, this differential equation needs to be (numerically) solved subject to the boundary condition
\begin{equation}
\lim_{t \to 0^+} \sigma(t) = -\frac{t}{\pi} - \bigg( \frac{t}{\pi} \bigg)^2.
\end{equation}
The corresponding boundary condition for $\sigma'(t)$ follows by taking the derivative. Setting the boundary at $t = 0$ has to be avoided because of the division by $t$. See Ref.~\cite{Edelman05} for a plot of the solution $\sigma(t)$.

For Dyson index $\beta$ (introduced above), the level spacing distribution $P_\beta(s)$ is here expressed in terms of a second derivative as
\begin{equation}
P_\beta(s) = \frac{d^2}{d s^2} E_\beta(s).
\label{eq: Ps-Painleve}
\end{equation}
The function $E_\beta(s)$ gives the probability not to find an eigenvalue in an arbitrary interval of width $s$. For unitary random matrices ($\beta = 2$), $E_\beta(s)$ is given by
\begin{equation}
E_2(s) = \exp \bigg( \int_0^{\pi s} \frac{\sigma(t)}{t} dt \bigg).
\label{eq: E2}
\end{equation}
Substituting Eq.~\eqref{eq: E2} in Eq.~\eqref{eq: Ps-Painleve} and evaluating the derivative explicitly leads to
\begin{equation}
P_2 (s) = \frac{1}{s^2} \bigg( \pi s \, \sigma' (\pi s) - \sigma (\pi s) + \big(\sigma (\pi s) \big)^2 \bigg) E_2(s).
\end{equation}
With the solution $\sigma(t)$, $\sigma'(t)$ at hand, this expression can be evaluated without the need for numerical differentiation. This is advantageous from a numerical point of view, as the use of numerical differentiation might lead to a loss of accuracy. Reference~\cite{Forrester00} points out that for orthogonal ($\beta = 1$) and symplectic ($\beta = 4$) random matrices, one has
\begin{equation}
E_1(s) = \sqrt{E_2(s)} \, \exp \bigg[ -\frac{i}{2} \int_0^{\pi s} \bigg( \frac{d}{dt} \frac{\sigma(t)}{t} \bigg)^{1/2} dt \bigg]
\end{equation}
and
\begin{equation}
E_4(s / 2) = \frac{1}{2} \bigg( E_1(s) + \frac{E_2(s)}{E_1(s)} \bigg).
\end{equation}
See Refs.~\cite{Dyson62-2, Mehta63} for the works where these expressions were derived first. Explicit differentiation also here provides expressions that can be evaluated without the need for numerical differentiation. These expressions are not shown here because of their length and limited pedagogical value.

Numerically, Eq.~\eqref{eq: Painleve-V} is solved with the boundary condition fixed at $t = 10^{-14}$ \footnote{The differential equation is solved using the \texttt{NDSolve} function of Wolfram Inc. \emph{Mathematica} version 14.00 with options \texttt{Method -> \{"StiffnessSwitching", Method -> \{"ExplicitRungeKutta", Automatic\}\}}, \texttt{InterpolationOrder -> All}, \texttt{WorkingPrecision -> 50}, \texttt{MaxSteps -> Infinity}, and \texttt{MaxStepSize -> 0.001}. Level spacing distributions are evaluated with step sizes $0.0005$ ($\beta = 1$) or $0.00025$ ($\beta = 4$). Cubic spline interpolations are used to connect successive data points.}. Level spacing distributions are evaluated from $s = 0$ up to $s = 10$ ($\beta = 1$) or $s = 5$ ($\beta = 4$). Since $P_{\beta = 1}(s = 10) \approx 7.3 \times 10^{-29}$ and $P_{\beta = 4}(s = 5) \approx 5.3 \times 10^{-22}$, this truncation does not affect the results. Table~\ref{tab: normalization-mean} shows the normalization and mean level spacing for the resulting level spacing distributions as a benchmark. In all cases, the absolute deviation from the exact value, given by unity, is less than $10^{-14}$. A more careful numerical analysis of the problem could likely further improve the accuracy.

\begin{table}
\begin{tabular}{ l || l | l} 
$\beta$ 	& $\mathcal{N} $				& $\mu$ \\
\hline \hline
$1$		& $1.00000 \, 00000 \, 00006$	& $1.00000 \, 00000 \, 00001$ \\
$4$		& $1.00000 \, 00000 \, 00000$ 	& $1.00000 \, 00000 \, 00000$ \\
\end{tabular}
\caption{The normalization $\mathcal{N} = \int P_\beta(s) \, ds$ and mean level spacing $\mu = \int s \, P_\beta(s) \, ds$ for the level spacing distributions obtained numerically from Eq.~\eqref{eq: Ps-Painleve}. All absolute deviations from the exact values, given by unity, are less than $10^{-14}$.}
\label{tab: normalization-mean}
\end{table} 

Reference~\cite{Bornemann10} provides values for the variance, skewness, and kurtosis of the level spacing distributions considered here, obtained by a complementary numerical method. This Fredholm determinant method, in short, can be seen as an efficient way to obtain level spacing distributions through integrations over the (explicitly known) $n$-level correlation functions. See Ref.~\cite{Bornemann10-2} for a more detailed review. Table~\ref{tab: Bornemann} compares these statistics with the ones obtained here for each of the level spacing distributions as an additional benchmark. All values agree up to the uncertainty reported in the reference.

\begin{table}
\begin{tabular}{ l || l | l | l} 
$\beta$ 			& $\sigma^2$			& $\gamma_1$			& $\gamma_2$ \\
\hline \hline
$1$	(reference)	& $0.28553 \, 06557$	& $0.68718 \, 99889$	& $0.37123 \, 80638$ \\
$1$	(this work)	& $0.28553 \, 06557$	& $0.68718 \, 99889$	& $0.37123 \, 80639$ \\
$4$	(reference)	& $0.10409 \, 84222$	& $0.34939 \, 68438$	& $0.02858 \, 27332$ \\
$4$	(this work)	& $0.10409 \, 84222$	& $0.34939 \, 68439$	& $0.02858 \, 27333$ \\
\end{tabular}
\caption{The variance $\sigma^2 = \int (s - \mu)^2 \, P_\beta(s) \, ds$, skewness $\gamma_1 = (\sigma^2)^{-3/2} \int (x - \mu)^3 \, P_\beta(s) \, ds$, and kurtosis $\gamma_2 = (\sigma^2)^{-2} \int (x - \mu)^4 \, P_\beta(s) \, ds - 3$ for the level spacing distributions obtained numerically from Eq.~\eqref{eq: Ps-Painleve}. Reference values from Ref.~\cite{Bornemann10} are shown for comparison. All values agree up to the uncertainty reported in the reference.}
\label{tab: Bornemann}
\end{table}

\section{Finite-$\boldsymbol{N}$ average} \label{sec: finite-average}
Let $\langle q \rangle_N$ denote the average of $\langle q \rangle$ for orthogonal random matrices of dimension $N$. The rate of convergence of $\langle q \rangle_N$ towards $\langle q \rangle_\infty$ can be of interest when diagnosing ergodicity for quantum systems of small size. In order to avoid possible artefacts due to deviating statistics at the edges of the spectra, finite-dimensional random matrices are in this work sampled from the circular ensembles~\cite{Dyson62, Forrester10}. The eigenvalue density of these unitary (circular) matrices is uniform on the unit circle in the complex plane, meaning that there is no spectral edge.

Random matrices from the circular ensembles are here sampled as follows. A circular Dyson Brownian motion process evaluated at infinite time is used to generate a sample from the circular unitary ensemble first~\cite{Buijsman24}. For this, a singular value decomposition is performed on an $N \times N$ complex matrix $M$ sampled from the Ginibre unitary ensemble ($\text{GinUE}_N$),
\begin{equation}
M = U_1 \Sigma U_2^\dagger, \qquad M \in \text{GinUE}_N.
\end{equation}
A matrix from the Ginibre unitary ensemble has elements for which the real and imaginary parts are sampled independently from a Gaussian distribution with mean zero and unit variance. An $N$-dimensional sample from the circular unitary ensemble ($\text{CUE}_N$) is then obtained as
\begin{equation}
U = U_1 U_2^\dagger \stackrel{\text{d}}{=} \text{CUE}_N,
\end{equation}
where the symbol $\stackrel{\text{d}}{=}$ means equal in distribution. References~\cite{Eaton07, Mezzadri07} discuss an alternative algorithm of similar computational complexity, which is based on the singular value decomposition of $M$. An $N$-dimensional sample from the circular orthogonal ensemble ($\text{COE}_N$) is subsequently obtained through the relation
\begin{equation}
U^T U \stackrel{\text{d}}{=} \text{COE}_N
\end{equation}
\cite{Eaton07, Mezzadri07}. Eigenvalues of the circular symplectic ensemble can be sampled through the aforementioned inter-relation between the spectral statistics of orthogonal and symplectic random matrices, which for the circular ensembles can be shown relatively straightforward to hold also at finite dimension~\cite{Mehta63}. 

\begin{figure}
\includegraphics[scale=0.9]{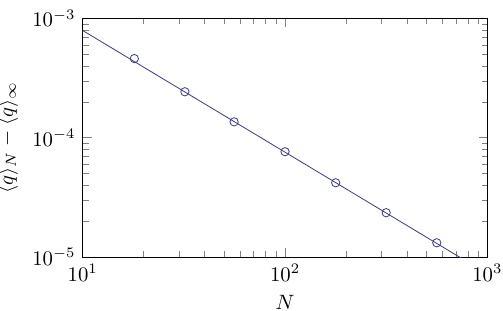}
\caption{The deviation of $\langle q \rangle_N$ from $\langle q \rangle_\infty$ versus $N$. The data points show numerically obtained averages over the spectra of $10^9$ matrices sampled from the circular orthogonal ensemble. For all data points, the sample standard deviation is less than $0.025$ times the sample mean. The line shows a fitted power law with coefficient $-1.02$. This coefficient has been obtained from a linear least-squares fit on the logarithms of the deviations versus $\ln(N)$. The first data point ($N = 18$) has been omitted in the fitting procedure.}
\label{fig: convergence}
\end{figure}

Fig.~\ref{fig: convergence} investigates the finite-size dependence of $\langle q \rangle_N$ by showing the deviation from $\langle q \rangle_\infty$ as a function of $N$. The data indicates a power-law decay, where the fitted power law coefficient is given by $-1.02$. Plausibly, the dependence follows a power-law decay with power law coefficient $-1$, with the deviation from the fitted value caused by finite-size effects. In order to find $\langle q \rangle_\infty$ at a precision of $10^{-10}$ (as in this work) from the spectra of finite-dimensional random matrices, one needs to consider matrices of dimension $\sim 10^8$. Numerical diagonalization of such high-dimensional matrices is challenging.

\section{Conclusions and outlook} \label{sec: conclusions}
A Painlev\'e differential equation has been solved numerically in order to compute the average of $q = \ln(1 + r)$ for infinite-dimensional orthogonal random matrices, where $r$ is the ratio of consecutive level spacings. This average $\langle q \rangle \approx 0.8100699350$ has been determined without the need for spectral unfolding. For Poissonian level statistics, a significantly different average $\langle q \rangle = 1$ has been found, indicating that $\langle q \rangle$ can be used as a probe for quantum ergodicity. Wigner surmise-inspired calculations for low-dimensional random matrices have next been shown to capture the distributions of $q$ for high-dimensional random matrices from each of the symmetry classes (orthogonal, unitary, symplectic) at a qualitatively accurate level. The results presented in this work could be useful in detecting small deviations from quantum ergodicity, and shows a way of how level spacing distributions can be used to study ratios of consecutive level spacings.

In future studies, it could be worth exploring the possibility to find the value of $\langle q \rangle$ for unitary and symplectic infinite-dimensional random matrices. Fredholm determinant methods, although being technically more involved than the method used here, allow for such computations~\cite{Bornemann10, Bornemann10-2}. These results would allow one to adapt $\langle q \rangle$ as a generically applicable probe for quantum ergodicity. Next, it might be interesting to determine $\langle q \rangle$ for the level statistics observed near the edges of the spectra of high-dimensional random matrices. Near the edges, level spacing statistics of high-dimensional random matrices obey a type of universality that is different from the one observed away from the edges, as studied in this work~\cite{Tracy94, Deift07}. Fredholm determinant methods also allow for the computation of $\langle q \rangle$ for these statistics~\cite{Bornemann10, Bornemann10-2}. L\'evy random matrix models might be interesting objects to study when focusing on spectral edge statistics, as orthogonal polynomial-based methods have lead to the suggestion these show non-standard, possibly universal characteristics for this type of matrix models~\cite{Burda02, Choi10, Buijsman23}.

\section*{Data availability}
The data that support the findings of this article are openly available~\cite{Zenodo}.

\appendix*
\section{$\mathbf{\langle q \rangle_r}$ for three-dimensional Gaussian unitary and symplectic random matrices}
This Appendix gives the Wigner surmise-inspired average weighted ratio of consecutive level spacings $\langle q \rangle_r$ for unitary and symplectic random matrices, obtained by substituting Eq.~\eqref{eq: Wigner-surmise-inspired} in Eq.~\eqref{eq: q-ratio} and evaluating the integral. For unitary random matrices, one finds
\begin{multline}
\langle q \rangle_r (\beta = 2) = -\frac{3 \sqrt{3}}{8 \pi} + \frac{3i}{2 \pi} \bigg[ \Li_2 \bigg( \frac{1}{2} - \frac{i \sqrt{3}}{2} \bigg) \\ - \Li_2 \bigg( \frac{1}{2} + \frac{i \sqrt{3}}{2} \bigg) \bigg].
\end{multline}
For symplectic random matrices, one finds
\begin{multline}
\langle q \rangle_r (\beta = 4) = -\frac{69 \sqrt{3}}{160 \pi} + \frac{3i}{2 \pi} \bigg[ \Li_2 \bigg( \frac{1}{2} - \frac{i \sqrt{3}}{2} \bigg) \\ - \Li_2 \bigg( \frac{1}{2} + \frac{i \sqrt{3}}{2} \bigg) \bigg].
\end{multline}
In these expressions, the dilogarithm $\Li_2(z)$ is defined in the conventional way as
\begin{equation}
\Li_2(z) = - \int_0^z \frac{\ln(1 - u)}{u} du.
\end{equation}

\bibliography{references}

\end{document}